\documentclass[12pt]{article}

\usepackage[margin=1in]{geometry}

\usepackage{setspace}

\usepackage{graphicx}

\usepackage{enumitem}

\usepackage{microtype}
\usepackage{subfigure}
\usepackage{booktabs}
\usepackage{comment}
\usepackage{wrapfig}
\usepackage{arydshln}
\usepackage{enumitem}
\usepackage{multirow}
\usepackage{url}
\usepackage{natbib}
\usepackage{authblk}

\RequirePackage[colorlinks,citecolor=blue,linkcolor=blue,urlcolor=blue,pagebackref]{hyperref}

\usepackage{amsmath}
\usepackage{amssymb}
\usepackage{mathtools}
\usepackage{amsfonts}
\usepackage{bm}
\usepackage{bbm}

\usepackage{algorithm}
\usepackage{algorithmic}

\def\eqref#1{equation~\ref{#1}}

\def\1{\bm{1}}

\DeclareMathAlphabet{\mathsfit}{\encodingdefault}{\sfdefault}{m}{sl}
\SetMathAlphabet{\mathsfit}{bold}{\encodingdefault}{\sfdefault}{bx}{n}

\def\calR{{\mathcal{R}}}

\def\calW{{\mathcal{W}}}
\def\calX{{\mathcal{X}}}
\def\calY{{\mathcal{Y}}}

\def\bbE{{\mathbb{E}}}

\def\bbR{{\mathbb{R}}}

\newcommand{\E}{\mathbb{E}}

\newcommand{\R}{\mathbb{R}}

\DeclareMathOperator*{\argmin}{arg\,min}

\newcommand{\p}[1]{\left(#1\right)}
\newcommand{\sqb}[1]{\left[#1\right]}
\newcommand{\cb}[1]{\left\{#1\right\}}

\newcommand{\bigp}[1]{\big(#1\big)}

\newcommand{\Bigp}[1]{\Big(#1\Big)}

\newcommand{\Exp}[1]{\mathbb{E}\left[#1\right]}

\usepackage{amsthm}

\theoremstyle{plain}

\newtheorem{theorem}{Theorem}[section]

\usepackage[textsize=tiny]{todonotes}
\usepackage{multirow}
\usepackage{wrapfig}
\usepackage{subfigure}

\renewcommand{\eqref}[1]{(\ref{#1})}

\usepackage{xcolor}
\newcount\Comments  
\newcommand{\kibitz}[2]{\ifnum\Comments=1\textcolor{#1}{#2}\fi}

\usepackage[capitalize,noabbrev]{cleveref}

\allowdisplaybreaks

\title{Nearest Neighbor Matching as Least Squares Density Ratio Estimation and Riesz Regression}

\author{Masahiro Kato\thanks{Email: \texttt{mkato-csecon@g.ecc.u-tokyo.ac.jp}}$\,$}

\affil{The University of Tokyo}

\date{\today}

\begin{document}

\maketitle

\begin{abstract}
This study proves that Nearest Neighbor (NN) matching can be interpreted as an instance of Riesz regression for automatic debiased machine learning. \citet{Lin2023estimationbased} shows that NN matching is an instance of density-ratio estimation with their new density-ratio estimator. \citet{Chernozhukov2024automaticdebiased} develops Riesz regression for automatic debiased machine learning, which directly estimates the Riesz representer (or equivalently, the bias-correction term) by minimizing the mean squared error. In this study, we first prove that the density-ratio estimation method proposed in \citet{Lin2023estimationbased} is essentially equivalent to Least-Squares Importance Fitting (LSIF) proposed in \citet{Kanamori2009aleastsquares} for direct density-ratio estimation. Furthermore, we derive Riesz regression using the LSIF framework. Based on these results, we derive NN matching from Riesz regression. This study is based on our work \citet{Kato2025directbias} and \citet{Kato2025directdebiased}.
\end{abstract}

\section{Introduction}
Nearest-neighbor matching is a popular method for treatment effect estimation from observational data. \citet{Lin2023estimationbased} shows that nearest-neighbor (NN) matching can be derived from density ratio estimation (DRE), providing a novel density ratio estimator.

In this study, we show that the density ratio estimator proposed in \citet{Lin2023estimationbased} is equivalent to Least-Squares Importance Fitting (LSIF), proposed by \citet{Kanamori2009aleastsquares}, with a specific kernel function. Furthermore, we show that nearest-neighbor matching can be viewed as a specific implementation of Riesz regression, proposed by \citet{Chernozhukov2024automaticdebiased}, by demonstrating the equivalence between LSIF and Riesz regression. Our results imply the following: (i) we can directly apply various results established for LSIF to NN matching, such as the choice of alternative kernels, convergence rates, and possible extensions; (ii) we confirm the generality of LSIF and Riesz regression for a variety of problems by identifying their equivalence with NN matching.

In Section~\ref{sec:problemformulation}, we formulate our problem. In Section~\ref{sec:recaplin}, we review the arguments in \citet{Lin2023estimationbased}. In Section~\ref{sec:linlsif}, we show that the DRE method proposed in \citet{Lin2023estimationbased} is equivalent to LSIF, proposed in \citet{Kanamori2009aleastsquares}. In Section~\ref{sec:equivnnriesz}, we show that NN matching is equivalent to Riesz regression, proposed in \citet{Chernozhukov2024automaticdebiased}, by establishing the equivalence between Riesz regression and LSIF. The results in this study are partially based on our previous work \citet{Kato2025directbias} and \citet{Kato2025directdebiased}.

\section{Problem Formulation}
\label{sec:problemformulation}
Let $X \in \calX \subseteq \bbR^d$ be $d$-dimensional covariates with a covariate space $\calX$, $D \in \{1, 0\}$ be a treatment indicator, and $Y_i \in \calY \subseteq \bbR$ be an outcome with an outcome space $\calY$. Let $Y(1)$ and $Y(0)$ be the potential outcomes of treated and control units, where $Y = D \cdot Y(1) + (1 - D) \cdot Y(0)$ holds. We define the Average Treatment Effect (ATE) as 
\[\tau_0 \coloneqq \Exp{Y(1) - Y(0)}.\]
We observe samples $\{(X_i, D_i, Y_i)\}^n_{i=1}$, where $(X_i, D_i, Y_i)$ is an i.i.d. copy of $(X, D, Y)$. Our goal is to estimate $\tau_0$ using these observations.

\paragraph{Assumptions and notation.} This section describes the assumptions and notation. We assume that the covariates in the treated and control groups have densities. Let $f(x \mid D = 1)$ denote the probability density of covariates in the treated group, $f(x \mid D = 0)$ the probability density in the control group, and $f(x)$ the marginal covariate density. We also denote the joint density of $(X, D)$ by $f(x, d)$. Let us define the propensity score as $e_0(X) \coloneqq P(D = 1\mid X) = \frac{f(x, 1)}{f(x)}$ and assume that there exists a constant $\epsilon \in (0, 1/2)$ independent of $n$ such that $\epsilon < e_0(X) < 1 - \epsilon$. We assume that $Y(1)$ and $Y(0)$ are independent of $D$ conditional on $X$ (unconfoundedness). We also assume that $X$, $Y(1)$, and $Y(0)$ are bounded. These assumptions may be relaxed, but we adopt them for simplicity without loss of generality in our claims. We write $\mathbbm{1}(\cdot)$ for the indicator function. Let $n_1 = \sum^n_{i=1}\mathbbm{1}(D_i = 1)$ and $n_0 = \sum^n_{i=1}\mathbbm{1}(D_i = 0)$.

\section{Recap of \citet{Lin2023estimationbased} }
\label{sec:recaplin}

\subsection{DRE method in \citet{Lin2023estimationbased}}
\label{sec:linder}
\citet{Lin2023estimationbased} first proposes their own density ratio estimation method. Let $X \sim \nu_0$ and $Z \sim \nu_1$ be independent with absolutely continuous laws on $\R^d$. We observe i.i.d. samples $\{X_i\}_{i=1}^{N_0}$ and $\{Z_j\}_{j=1}^{N_1}$ and aim to estimate the density ratio
\[
r_0(x) \coloneqq \frac{f_{(1)}(x)}{f_{(0)}(x)},
\]
where $f_{(1)}$ and $f_{(0)}$ are the probability densities of $\nu_1$ and $\nu_0$, respectively. For $M \in \{1,\dots,N_0\}$ and $z \in \R^d$, let $\calX_{(M)}(z)$ be the $M$-th nearest neighbor of $z$ in $\{X_i\}_{i=1}^{N_0}$ under a given metric $\|\cdot\|$. Define the \emph{catchment area} of $x$ as
\[
A_M(x) \coloneqq \bigl\{z : \|x - z\| \le \|\calX_{(M)}(z) - z\|\bigr\},
\]
and the \emph{matched-times count} as
\[
K_M(x) \coloneqq \sum_{j=1}^{N_1} \mathbbm{1}\bigp{Z_j \in A_M(x)}.
\]
\citet{Lin2023estimationbased} propose the one-step estimator
\[
\widehat r_M(x) = \frac{N_0}{N_1} \frac{K_M(x)}{M},
\]
and show that it is computationally efficient and rate-optimal for Lipschitz densities.

\subsection{NN matching}
The $M$–NN matching estimator of \citet{Abadie2006largesample} imputes the missing potential outcome of unit $i$ by averaging the observed outcomes of its $M$ nearest neighbors in the opposite treatment arm. Let
\[
J_M(i)\ \subset\ \{1,\ldots,n\}
\]
be the index set of the $M$ nearest neighbors of $X_i$ among the units with $D_j = 1 - D_i$. The imputed outcomes are
\[
\widehat Y_i(0) =
\begin{cases}
Y_i, & D_i = 0,\\[2mm]
\dfrac{1}{M} \sum_{j \in J_M(i)} Y_j, & D_i = 1,
\end{cases}
\qquad
\widehat Y_i(1) =
\begin{cases}
\dfrac{1}{M} \sum_{j \in J_M(i)} Y_j, & D_i = 0,\\[2mm]
Y_i, & D_i = 1.
\end{cases}
\]
The NN matching estimator for the average treatment effect (ATE) is
\[
\widehat\tau_M \coloneqq \frac{1}{n} \sum_{i=1}^n \p{\widehat Y_i(1) - \widehat Y_i(0)}.
\]
Introduce the \emph{matched-times count} (how many times unit $i$ is used as a match by units in the opposite group) as
\[
K_M(i) \coloneqq \sum_{j=1,\ D_j = 1 - D_i}^n \mathbbm{1}\Bigp{i \in J_M(j)}.
\]
Then, $\widehat\tau_M$ can be written as follows:
\begin{equation}
\label{eq:ai06-linear}
\widehat\tau_M = \frac{1}{n} \sqb{ \sum_{i : D_i = 1} \Bigl(1 + \frac{K_M(i)}{M} \Bigr) Y_i - \sum_{i : D_i = 0} \p{1 + \frac{K_M(i)}{M}} Y_i }
= \frac{1}{n} \sum_{i=1}^n (2D_i - 1) \p{1 + \frac{K_M(i)}{M}} Y_i.
\end{equation}

\paragraph{Connection to DRE.}
In ATE estimation, the density-ratio estimator proposed in \citet{Lin2023estimationbased} is given as
\[
\frac{n_0}{n_1} \frac{K_M(i)}{M} \xrightarrow{p} \frac{f(X_i\mid D_i = 1)}{f(X_i \mid D_i = 0)} \quad (D_i = 0),
\qquad
\frac{n_1}{n_0} \frac{K_M(i)}{M} \xrightarrow{p} \frac{f(X_i\mid D_i = 0)}{f(X_i\mid D_i = 1)} \quad (D_i = 1),
\]
so that, using $n_1 / n_0 \to P(D = 1) / P(D = 0)$, the factor $1 + \frac{K_M(i)}{M}$ is a consistent estimator of the inverse propensity weights:
\[
1 + \frac{K_M(i)}{M} \xrightarrow{p}
\begin{cases}
\dfrac{1}{1 - e_0(X_i)}, & D_i = 0,\\[2mm]
\dfrac{1}{e_0(X_i)}, & D_i = 1.
\end{cases}
\]
Substituting these weights into \eqref{eq:ai06-linear} gives the DRE–motivated representation of NN matching.

\paragraph{Bias-corrected NN matching.}
Let $\widehat\mu_\omega(x)$ estimate $\mu_\omega(x) \coloneqq \E[Y \mid X = x, D = \omega]$, and set residuals $\widehat R_i \coloneqq Y_i - \widehat\mu_{D_i}(X_i)$ as well as the regression plug-in $\widehat\tau_{\mathrm{reg}} \coloneqq \frac{1}{n} \sum_{i=1}^n \bigl\{\widehat\mu_1(X_i) - \widehat\mu_0(X_i)\bigr\}$. The bias-corrected $M$–NN estimator of \citet{Abadie2011biascorrected} can be written as
\begin{equation}
\label{eq:bc-lin}
\widehat\tau_{bc,M}
= \widehat\tau_{\mathrm{reg}}
+ \frac{1}{n} \sqb{ \sum_{i : D_i = 1} \p{1 + \frac{K_M(i)}{M}} \widehat R_i - \sum_{i : D_i = 0} \p{1 + \frac{K_M(i)}{M}} \widehat R_i}.
\end{equation}
With $M \to \infty$ and $M \log n / n \to 0$, the weights $1 + \frac{K_M(i)}{M}$ consistently estimate the inverse propensity scores, so $\widehat\tau_{bc,M}$ is doubly robust and achieves the semiparametric efficiency bound under suitable smoothness and overlap conditions \citep{Lin2023estimationbased}.

\section{LSIF}
\label{sec:lsif}
LSIF aims to solve the least-squares problem
\[
r^* \coloneqq \argmin_{r \in \calR} \bbE_{f_{(0)}} \big[\bigp{r_0(X) - r(X)}^2\big],
\]
where $\calR$ denotes a hypothesis class for $r_0$, and $\bbE_{f_{(0)}}$ denotes expectation with respect to the density $f_{(0)}(x)$. Using the identity
\[
\bbE_{f_{(0)}} \big[(r_0 - r)^2\big]
= \bbE_{f_{(0)}} \big[r(X)^2\big] - 2\bbE_{f_{(1)}} \big[r(X)\big] + \text{const},
\]
we obtain the equivalent optimization that does not involve $r_0$:
\[
r^* = \argmin_{r \in \calR} \Big\{\bbE_{f_{(0)}} \big[r(X)^2\big] - 2\bbE_{f_{(1)}} \big[r(X)\big]\Big\},
\]
where the constant term is irrelevant for optimization.

We consider a linear-in-parameters density-ratio model; that is, with a basis $\Phi(x) \in \R^d$ we approximate $r_0$ as
\[
r_\beta(x) = \beta^\top \Phi(x),
\]
where $\beta \in \R^d$. We estimate $\beta$ by ridge regression:
\[
\widehat{\beta} \coloneqq \argmin_{\beta \in \bbR^d}
\cb{
\frac{1}{N_0} \sum_{i=1}^{N_0} r_\beta(X_i)^2
- 2\frac{1}{N_1} \sum_{j=1}^{N_1} r_\beta(Z_j)
+ \frac{\lambda}{2} \|\beta\|_2^2
},
\]
where $\lambda \ge 0$ is a regularization parameter, $\{X_i\}_{i=1}^{N_0} \sim f_{(0)}$, and $\{Z_j\}_{j=1}^{N_1} \sim f_{(1)}$.

This program has the closed-form solution
\[
\widehat\beta
= \argmin_{\beta \in \R^d}
\cb{\frac{1}{2} \beta^\top \widehat H \beta - \beta^\top \widehat h + \frac{\lambda}{2} \|\beta\|_2^2}
= \p{\widehat H + \lambda I}^{-1} \widehat h,
\]
where 
\begin{align*}
\widehat H &\coloneqq \frac{1}{N_0} \sum_{i=1}^{N_0} \Phi(X_i)\Phi(X_i)^\top,\\
\widehat h &\coloneqq \frac{1}{N_1} \sum_{j=1}^{N_1} \Phi(Z_j).
\end{align*}

\section{Equivalence between \citet{Lin2023estimationbased}'s DRE Method and LSIF}
\label{sec:linlsif}
We continue considering the DRE problem in Section~\ref{sec:linder}. We show the equivalence between \citet{Lin2023estimationbased}'s DRE method and LSIF, which also implies that NN matching can be justified via LSIF.

We aim to estimate $r_0(c)$ for a fixed point of interest $c \in \{Z_j\}^{N_1}_{j=1}$. Let us define the following $1$-dimensional basis function:
\[\Phi(x) = \Phi_c(x) = \mathbbm{1}\bigp{x \in A_M(c)},\]
where recall that $A_M(x) \coloneqq \bigl\{z : \|x - z\| \le \|\calX_{(M)}(z) - z\|\bigr\}$, and $\calX_{(M)}(z)$ is the $M$-th nearest neighbor of $z$ in $\{X_i\}_{i=1}^{N_0}$ under a given metric $\|\cdot\|$.

Then, we define a linear-in-parameters density-ratio model as
\[\widehat{r}(c) = \beta \Phi_c(c),\]
where $\beta$ is a scalar. Here, recall that LSIF gives 
\[\widehat\beta
= \p{\widehat H + \lambda I}^{-1} \widehat h,\]
where 
\begin{align*}
&\widehat H \coloneqq \frac{1}{N_0} \sum_{i=1}^{N_0} \Phi_c(X_i)^2 = \frac{1}{N_0} \sum_{i=1}^{N_0} \Phi_c(X_i) = \frac{M}{N_0},\\
&\widehat h \coloneqq \frac{1}{N_1} \sum_{j=1}^{N_1} \Phi_c(Z_j) = \frac{K_M(c)}{N_1},\\
&\Phi_c(c) = 1.
\end{align*}
Recall that $K_M(c) \coloneqq \sum_{j=1}^{N_1} \mathbbm{1}\bigp{Z_j \in A_M(c)}$.
Therefore, when $\lambda = 0$, we have
\[\widehat\beta
= \p{\widehat H + \lambda I}^{-1} \widehat h = \frac{N_0}{N_1} \cdot \frac{K_M(c)}{M},\]
and $\widehat{r}(c) = \widehat{\beta} \Phi_c(c) = \frac{N_0}{N_1} \cdot \frac{K_M(c)}{M}$.  
This estimator coincides with $\widehat r_M(c)$ derived in \citet{Lin2023estimationbased}. Thus, we show that the DRE method in \citet{Lin2023estimationbased} is equivalent to LSIF with the basis function $\widehat{r}(c) = \beta \Phi_c(c)$.

\begin{theorem}[LSIF reproduces $\widehat r_M$]\label{lem:lsif-nn}
With a basis function defined as $\Phi(x) = \mathbbm{1}\bigp{\|x - c\| \leq \|\calX_{(M)}(c) - c\|}$ and $\lambda = 0$, the LSIF estimator of $r_0(c)$ equals
\[
\widehat r(c) = \frac{N_0}{N_1} \cdot \frac{K_M(c)}{M} = \widehat r_M(c).
\]
\end{theorem}

Theorem~\ref{lem:lsif-nn} shows that the density-ratio estimator of \citet{Lin2023estimationbased} is an LSIF solution for a particular basis function. Hence, general properties of LSIF provide an alternative route to their results. In particular, we can interpret the density-ratio estimator of \citet{Lin2023estimationbased} as LSIF with a specific kernel function. 

\paragraph{Connection to existing DRE methods.}
This equivalence result implies that we can leverage various results established for DRE. In particular, \citet{Sugiyama2011densityratio} shows that a class of DRE methods is equivalent from the viewpoint of Bregman divergence. This class includes Kullback–Leibler divergence minimization \citep{Nguyen2007estimatingdivergence,Sugiyama2008directimportance} and learning from positive and unlabeled data \citep{Lancaster1996casecontrolstudies,duPlessis2015convexformulation,Kato2019learningfrom}. For example, we can apply convergence rate results shown in \citet{Kanamori2012statisticalanalysis} for more general reproducing kernel Hilbert spaces (RKHS). We can also incorporate the covariate shift setting, as investigated in \citet{Uehara2020offpolicy}, \citet{Kato2024doublyrobust}, and \citet{Chernozhukov2025automaticdebiased}. When applying neural networks, overfitting can be prevented by using regularization techniques such as those proposed in \citet{Rhodes2020} and \citet{Kato2021nonnegativebregman}.

Note that, since $\calR$ is restricted to $\{\beta\Phi_c(x)\}$, for 
$r^* \coloneqq r^*_c \coloneqq \argmin_{r_c \in \{\beta\Phi_c(x)\}} \bbE_{f_{(0)}} \big[\bigp{r_0(X) - r_c(X)}^2\big]$,
we usually have $r^*_c \neq r_0$. In contrast, existing studies analyze estimators under the assumption that $r_0 \in \calR$, as shown in \citet{Kanamori2012statisticalanalysis}. Even when $r_0 \notin \calR$, in many cases, we can still apply existing results to obtain a bound on the mean squared error (MSE). For example, by applying Theorem~2 in \citet{Kanamori2012statisticalanalysis}, we can bound $\Exp{\bigp{r^*_c(X) - \widehat{r}_c(X)}^2}$ for $\widehat{r}_c = \widehat{\beta}\Phi_c(x)$. In addition, we can assess how large a sample size is needed to achieve $r^*_c(c) \approx r_0(c)$.

We can also derive an MSE bound for $r_0$ over the entire $x \in \calX$ by considering a hypothetical estimator $\widetilde{r}(x)$ such that, for each $x$, we use $\widehat{r}_x(x) = \widehat{\beta}\Phi_x(x)$ obtained via NN matching. Then, the MSE can be bounded as $\Exp{\bigp{r_0(X) - \widetilde{r}(X)}^2} = 2\p{\Exp{\bigp{r_0(X) - r^*_X(X)}^2} + \Exp{\bigp{r^*_X(X) - \widehat{r}_X(X)}^2}}$, which can be evaluated via the existing results.

\section{Equivalence among NN Matching and Riesz Regression}
\label{sec:equivnnriesz}
This section further shows the equivalence among nearest-neighbor matching, LSIF, and Riesz regression. To demonstrate this, we show that the bias-correction term can be estimated using LSIF in a more direct way than the formulation of \citet{Lin2023estimationbased}. That is, we can directly estimate the inverse propensity scores by interpreting them as density ratios:
\[
w_0(1, X) = \frac{1}{e_0(X)} = \frac{f(X)}{f(X, 1)},\qquad w_0(0, X) \coloneqq \frac{1}{1 - e_0(X)} = \frac{f(X)}{f(X, 0)}.
\]
Note that the LSIF formulation in this section is nearly the same as the one in Section~\ref{sec:lsif}, and corresponds to the expressions involving $1 + \frac{K_M(i)}{M}$ in \citet{Lin2023estimationbased}. We consider this formulation more straightforward than that in \citet{Lin2023estimationbased}.

\subsection{Another Formulation of LSIF}
We first consider LSIF to estimate $w_0(1, \cdot)$. Note that we separately estimate $w_0(1, \cdot)$ and $w_0(0, \cdot)$, and $w_0(0, \cdot)$ can be estimated in a similar way.

We aim to estimate $w_0(1, c)$ for a fixed point of interest $c = X_i$ such that $D_i = 1$. Following LSIF, we can estimate an empirical version of the following least-squares problem:
\begin{align}
\label{eq:lsifimprove1}
    w^*(1, \cdot) &\coloneqq \argmin_{w \in \calW} \bbE_{f(X, 1)}\sqb{\bigp{w_0(1, X) - w(1, X)}^2}\\
    &= \argmin_{w \in \calW} \Big\{\bbE_{f(X, 1)}[w(1, X)^2] - 2 \bbE_{f(X)}[w(1, X)]\Big\},
\end{align}
where the expectation is taken under the denominator $f(X, 1)$ for the squared term and the numerator $f(X)$ for the linear term (dropping constants).

Since our interest lies only in the point $c$, we use the following one-dimensional basis function again:
\[
\Phi(x) = \Phi_c(x) = \mathbbm{1}\bigp{x \in A_M(c)}.
\]

We consider LSIF with a linear-in-parameter model
\[
w(1, c) = \theta \Phi_c(c),
\]
where $\theta$ is a scalar. Then, the estimator is given as
\[
\widehat w(1, c) = \widehat\theta^\top \Phi(c)
\]
with the estimated parameter
\[
\widehat\theta \coloneqq \argmin_{\theta \in \R}
\cb{\frac{1}{2} \theta^\top \widehat H \theta - \theta^\top \widehat h + \frac{\lambda}{2} \|\theta\|_2^2}
= \p{\widehat H + \lambda I}^{-1} \widehat h,
\]
where
\begin{align*}
\widehat H &\coloneqq \frac{1}{n} \sum_{i=1}^{n} \mathbbm{1}(D_i = 1) \Phi_c(X_i)^2,\\
\widehat h &\coloneqq \frac{1}{n} \sum_{i=1}^{n} \Phi_c(X_i).
\end{align*}

Here, we have
\begin{align*}
\widehat H &= \frac{1}{n} \sum_{i=1}^{n} \mathbbm{1}(D_i = 1) \Phi_c(X_i) = \frac{M}{n},\\
\widehat h &= \frac{1}{n} \sum_{i=1}^{n} \Phi_c(X_i) = \frac{1}{n} \sum_{i=1}^{n} \bigp{\mathbbm{1}(D_i = 1) \Phi_c(X_i) + \mathbbm{1}(D_i = 0) \Phi_c(X_i)} = \frac{1}{n} \bigp{M + K_M(i)},\\
\Phi_c(c) &= 1,
\end{align*}
where recall that
\[
K_M(i) \coloneqq \sum_{j=1,\ D_j = 1 - D_i}^n \mathbbm{1}\Bigp{i \in J_M(j)}.
\]

Therefore, when $\lambda = 0$, we estimate $\widehat w(1, x)$ as
\[
\widehat w(1, x) = 1 + \frac{K_M(i)}{M}.
\]
Similarly, we can estimate $\widehat w(0, c)$  for a fixed point of interest $c = X_i$ such that $D_i = 0$ by solving an empirical version of the following problem:
\begin{align}
\label{eq:lsifimprove2}
    w^*(0, \cdot) &\coloneqq \argmin_{w \in \calW} \Big\{\bbE_{f(X, 0)}[w(0, X)^2] - 2 \bbE_{f(X)}[w(0, X)]\Big\}.
\end{align}
Then, the estimator is given as
\[
\widehat w(0, c) = 1 + \frac{K_M(i)}{M}.
\]
Using these estimators, we construct the following inverse propensity score estimator for the ATE:
\[
\widehat\tau_M = \frac{1}{n} \sqb{ \sum_{i : D_i = 1} \p{1 + \frac{K_M(i)}{M}} Y_i - \sum_{i : D_i = 0} \p{1 + \frac{K_M(i)}{M}} Y_i }.
\]
This estimator is equivalent to \eqref{eq:ai06-linear} proposed in \citet{Lin2023estimationbased}, which is shown to be equal to the NN matching estimator of \citet{Abadie2006largesample}. We can also construct the bias-corrected estimator in a similar way. Thus, we have shown the equivalence between LSIF, the method in \citet{Lin2023estimationbased}, and NN matching in a different way from Section~\ref{sec:linlsif}.

\subsection{Riesz Regression}
In debiased machine learning, the estimation of the Riesz representer plays a core role \citep{Chernozhukov2022automaticdebiased}. To estimate the Riesz representer effectively, Riesz regression has been developed by \citet{Chernozhukov2024automaticdebiased}. In this context, Riesz regression is essentially the same as LSIF in \citet{Kanamori2009aleastsquares}.

Let $W = (Y, Z)$, where $Z$ is a regressor and $\gamma_0 \coloneqq \Exp{Y \mid Z}$ is a regression function. We consider the target parameter to be a linear functional of the regression function, $\tau_0 = \bbE[m(W, \gamma_0)]$, with Riesz representer $\alpha_0$ satisfying
\[
\bbE\bigl[m(W, \gamma)\bigr] = \bbE\bigl[\alpha_0(X) \gamma(X)\bigr] \quad \forall \gamma.
\]
\citet{Chernozhukov2024automaticdebiased} show that $\alpha_0$ is the minimizer of the \emph{Riesz regression} risk:
\[
\alpha_0 = \argmin_{\alpha} \bbE\left[\alpha(X)^2 - 2 m(W, \alpha)\right],
\]
leading to the orthogonal, doubly robust score
\[
\psi(W, \gamma, \alpha, \tau) = m(W, \gamma) - \tau + \alpha(Z)\bigl(Y - \gamma(Z)\bigr),
\]
whose expectation is zero if either $\gamma = \gamma_0$ or $\alpha = \alpha_0$.

In ATE estimation, for $Z = (D, X)$, the Riesz representer is given as
\[
\alpha_0(Z) = \mathbbm{1}(D = 1) w_0(1, X) - \mathbbm{1}(D = 0) w_0(0, X).
\]
For this Riesz representer, the Riesz regression objective is
\begin{align}
\label{eq:rieszobjective}
\alpha_0 &= \argmin_{\alpha} \bbE\left[\alpha(Z)^2 - 2 \alpha(Z)\right]\\
&= \argmin_{\alpha} \bbE\left[\mathbbm{1}(D = 1) w(1, X)^2 + \mathbbm{1}(D = 0) w(0, X)^2 - 2 \bigp{w(1, X) + w(0, X)}\right].
\end{align}

This objective function \eqref{eq:rieszobjective} for Riesz regression can also be obtained by simultaneously solving the problems \eqref{eq:lsifimprove1} and \eqref{eq:lsifimprove2}. Therefore, Riesz regression is equivalent to LSIF, which implies that NN matching is a specific case of Riesz regression.

\section{Discussion and implications}
Our results provide a constructive dictionary in which Riesz regression collapses to LSIF and to the NN density-ratio estimator, yielding an explicit bridge from automatic debiased machine learning to NN matching:
\[
\boxed{\text{NN matching}}\ \equiv\ \boxed{\text{Riesz regression}}\ \equiv\ \boxed{\text{LSIF}}.
\]
This equivalence clarifies why increasing $M$ improves statistical efficiency and why the same estimator inherits both graph-based computational advantages and LSIF’s quadratic-loss simplicity.

The results of this study are partially shown in \citet{Kato2025directbias} and \citet{Kato2025directdebiased}. In those studies, further extensions have been proposed. For example, they show that Riesz regression can be generalized by incorporating Bregman divergence, which also bridges covariate balancing and Riesz regression. In fact, the Riesz representer estimated via LSIF or Riesz regression is in a dual relationship with the stable weights of \citet{Zubizarreta2015stableweights}. In addition, they show the connection between Riesz regression and targeted maximum likelihood estimation (TMLE) in \citet{vanderLaan2006targetedmaximum}. Combining these findings with the results of this study, we can relate NN matching with LSIF, Riesz regression, covariate balancing, and TMLE. Furthermore, based on the results of \citet{BrunsSmith2025augmentedbalancing}, the bias-corrected NN estimator can be expressed as a single linear model.

\bibliography{arXiv2.bbl}

\bibliographystyle{tmlr}

\end{document}